\documentclass{piparticle-final}
\usepackage[utf8]{inputenc}
\usepackage{graphicx}
\usepackage{amsmath}

\usepackage{cite} 
\usepackage{epstopdf} 
\newcommand {\pri}{^{\prime }}

\author{\small{Miguel A. R\'e}\\ \scriptsize{Departamento de Ciencias Básicas, CIII, Facultad Regional Córdoba, Universidad Tecnol\'ogica Nacional} \\ \scriptsize{ Facultad de Matemática, Astronomía y Física, Universidad Nacional de Córdoba}\\ \scriptsize{Córdoba, Argentina} \\
\small{Natalia C. Bustos}\\ \scriptsize{ Facultad de Matemática, Astronomía y Física, Universidad Nacional de Córdoba}\\ \scriptsize{Córdoba, Argentina}}

\begin{document}

\volume{7}               
\articlenumber{070003}   
\journalyear{2015}       
\editor{C. A. Condat, G. J. Sibona}   
\received{20 November 2014}     
\accepted{20 March 2015}   
\runningauthor{M. A. R\'e \itshape{et al.}}  
\doi{070003}         

\title{Reaction rate in an evanescent random walkers system}

\author{Miguel A. R\'e,\cite{inst1,inst2}\thanks{Email: mgl.re33@gmail.com}\hspace{0.5em}
	Natalia C. Bustos\cite{inst2}
	}

\pipabstract{
Diffusion mediated reaction models are particularly ubiquitous in the description of physical, chemical or biological processes. The random walk schema is a useful tool for formulating these models. Recently, evanescent random walk models have received attention in order to include finite lifetime processes. For instance, activated chemical reactions, such as laser photolysis, exhibit a different asymptotic limit when compared with immortal walker models. A diffusion limited reaction model based on a one dimensional continuous time random walk on a lattice with evanescent walkers is presented here. The absorption probability density and the reaction rate are analytically calculated in the Laplace domain. A finite absorption rate is considered, a model usually referred to as imperfect trapping. Short and long time behaviors are analyzed.
}

\maketitle

\blfootnote{
\begin{theaffiliation}{99}
	\institution{inst1} Departamento de Ciencias B\'asicas, CIII, Facultad Regional C\'ordoba, Universidad Tecnol\'ogica Nacional. C\'ordoba, Argentina.
	\institution{inst2} Facultad de Matem\'atica, Astronom\'ia y F\'isica, Universidad Nacional de C\'ordoba, Ciudad Universitaria, 5010 C\'ordoba, Argentina.
\end{theaffiliation}
}

\section{Introduction}

The dynamics of the diffusion mediated reaction process has been extensively studied for many years due to its relevance in the description of diverse phenomena in physics, chemistry or biology \cite{weiss,rices,bergh,goeln}. A particularly interesting problem is the calculation of the probability density for the time at which a reaction $A+B\rightarrow C$ takes place (Absorption Probability Density - APD)  when the displacement of species $A$ or $B$ (or both) is diffusive. Other magnitudes, such as time dependent reaction rates or survival probabilities of the reactives, can be derived from the APD. Dielectric relaxation \cite{bendl}, capture of ligands after surface diffusion \cite{wangd} or proteins with active sites deep inside the protein matrix \cite{nadle} are examples of the application of the diffusion mediated reactions schema.\\
A random walk schema provides an excellent tool to model diffusion and has been studied for a long time with different alternatives to include the reaction process. Recently, a new kind of random walk models has been addressed in \cite{yuste,abadk} to include {\it evanescent} or {\it mortal} random walkers. In these models, the diffusing particles (reactives) may disappear during their displacement. Their disappearance may represent, for example, the decay of a laser activated reactive as in studies of fluorescence quenching by laser photolysis \cite{chuan}.\\
The calculation of chemical reaction rates from a model of {\it immortal} diffusing particles in the presence of a trap may be traced to the original contribution of Smoluchowski \cite{smolu}: a single absorbing sphere surrounded by diffusing particles with an initial uniform concentration. Smoluchowski's model assumes immediate trapping (reaction) upon encounter of the reactives, a great dilution of one of the species (the minority species represented by the sphere)  and normal diffusion of the particles with diffusion coefficient $D=D_A+D_B$, being $D_A$ and $D_B$ the diffusion coefficients of species $A$ and $B$, respectively. Diverse extensions have been proposed to include diffusion in disordered media \cite{bendl} or to improve the description of short time behavior by considering a finite reaction rate \cite{colli,noyes}. Extensions of Smoluchowski's model to random walks on lattices have been proposed to consider a finite reaction rate \cite{conda,rebud} or the 
modulation of reaction through gating 
controlled by an independent dynamics \cite{cacer}. In these models, the reactives may separate without reaction in each encounter.\\
We consider here an evanescent Continuous Time Random Walk (CTRW) on a one dimensional lattice with transitions to the nearest neighbors. We assume a time independent transitions rate $\lambda $ for diffusion and $\eta $ for evanescence. The reaction is included in the model by considering a trap at a fixed position in the lattice. When a walker arrives at this position, it may be trapped with a rate $\kappa $ or it may escape to a neighbor site with transition rate $\lambda $. The main magnitude to be calculated is the Absorption Probability Density (APD) for a walker starting at an arbitrary position on the lattice: the probability density for the time of reaction. The APD is analytically calculated in the Laplace representation and from this magnitude the Reaction Rate and the Survival Probability are calculated.

\section{Continuous time random walk}

\begin{figure*}[t]
\begin{center}
\setlength{\unitlength}{0.19cm}
\begin{picture}(80,15)(-20,5)
{\put(-20,10){\circle{1.5}}\put(-15,10){\circle{1.5}}\put(-10,10){\circle{1.5}}{\put(-7.5,10){\ldots}}}
\put(-14,10){\oval(2,2)[lt]}
\put(-14,12){\oval(2,2)[rb]}
\put(-14,12){\oval(2,2)[rt]}
\put(-14,14){\oval(2,2)[lb]}
\put(-15,14){\vector(0,1){1}}
\put(-15,10){\circle*{0.5}}
\put(-15,8.5){\makebox(0,0)[tc]{evanescence}}
\put(-15,12){\makebox(0,0)[cr]{$\eta $}}

{\put(0,10){\circle{1.5}}\put(5,10){\circle{1.5}}{\put(7.5,10){\ldots}}}
{\put(50,10){\circle{1.5}}\put(44,9){\framebox(2.0,2.0){}}\put(45,10){\circle*{0.5}}\put(45,10){\vector(0,1){5}}\put(44,14){\framebox(2.0,2.0){}}{\put(37.5,10){\ldots}}}
{\put(20,10){\circle{1.5}}\put(25,10){\circle{1.5}}\put(25,10){\circle*{0.5}}\put(30,10){\circle{1.5}}}
\put(22.5,10){\oval(5,5)[rt]}
\put(22.5,12.5){\vector(-1,0){3}}
\put(18,15){\makebox(0,0)[b]{$p_i\lambda $}}
\put(32,15){\makebox(0,0)[b]{$p_d\lambda $}}
\put(27.5,10){\oval(5,5)[lt]}
\put(27.5,12.5){\vector(1,0){3}}
\put(25,8.5){\makebox(0,0)[tc]{diffusion}}
\put(45,8.5){\makebox(0,0)[tc]{trapping}}
\put(43,15){\makebox(0,0)[cr]{limbo}}
\put(47,12.5){\makebox(0,0)[cl]{$\kappa $}}

\end{picture}

\end{center}
\caption{Evanescent CTRW on a one dimensional lattice. The walker may evanesce with an evanescent rate $\eta $. There is a trap at a particular site of the lattice. When the walker reaches the trap position, it may be trapped with a trapping rate $\kappa $.}
\label{fig1}
\end{figure*}
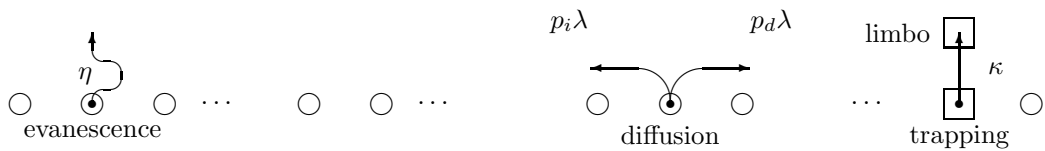

We include here some general CTRW results. Although most of these results may be found in the literature, we include them here with our particular problem in mind and also to make consistent the notation used in this paper.\\
Let us consider an infinite one dimensional lattice as shown in Fig. \ref{fig1}. Each position in the lattice is identified by an integer number $x$. We assume that at some instant $t=0$; present on the lattice, there is a uniform distribution of noninteracting walkers with concentration $c_0$ at every lattice position. Each walker is able to perform a CTRW with probability $\psi _0\left( x-x\pri ; t-t\pri  \right) dt$ of making a transition $x\pri \rightarrow x$ between $t$ and $t+dt$, having arrived at $x\pri  $ at time $t\pri  $. We shall assume here the waiting time probability density

\begin{equation}
\label{e1.1}
\psi _0\left( x-x\pri; t\right) = \left[ p_d\, \delta _{x,x\pri +1} + p_i\, \delta _{x,x\pri -1} \right] \lambda e^{-\lambda t}.
\end{equation}

Let $G_0\left( x;t\mid x_0\right) $ denote the conditional probability density for the arrival time at position $x$ of a walker that started its journey at $x_0$. This probability density is the Green's function for the problem and satisfies the recursive relation

\begin{align}
\label{e1.2}
G_0\left( x;t\mid x_0 \right) &= \delta _{x,x_0}\delta \left( t-0^+\right) \\
& + \sum \limits_{x\pri } \psi _0\left( x-x\pri ;t\right) \star G_0\left( x\pri ;t\mid x_0\right),  \notag
\end{align}
where $\star $ stands for the time convolution product

$$
f\left( t\right) \star g\left( t\right) = \int \limits _0^t \, dt\pri f\left( t-t\pri \right) g\left( t\pri \right). 
$$
Equation (\ref{e1.2}) may be solved by taking Laplace transform in the time variable and Fourier transform in the lattice coordinate. By means of this procedure, we get the solution in the Laplace representation 

\begin{equation}
\label{e1.3}
G^L_0\left( x;u\mid x_0\right)  = \dfrac{1}{2R_p\psi ^L_0\left( u\right) R\left( u\right) }\dfrac{\left[\xi \left( u\right) \right]^{\mid x-x_0\mid }}{R_c^{x-x_0}}
\end{equation}
where the super index $ ^L$ indicates the Laplace transform of a function

$$
f^L\left( u\right) = \int _0^{\infty }\, dt\, e^{-ut} \, f\left( t\right),
$$
and we have defined the auxiliary symbols

\begin{align}
\label{e1.4}
&R_p = \sqrt{ p_dp_i}, \ \ \  R_c = \sqrt{p_i/p_d}, \notag \\
&\psi ^L_0\left( u\right) = \dfrac{\lambda }{u+\lambda }, \ \ \  R\left( u\right) = \sqrt{ \left( \dfrac{1}{2R_p\psi ^L_0\left( u\right) }\right) ^2-1}, \notag \\
&\xi \left( u\right) = \dfrac{1}{2R_p\psi ^L_0\left( u\right) }- R\left( u\right).
\end{align}
The conditional probability $P_0\left( x;t\mid x_0\right) $ of finding the walker at position $x$ at time $t$ given that it started at $x_0$ can be obtained from the Green's function by a convolution product

\begin{equation}
\label{e1.5}
P_0\left( x;t\mid x_0\right) = \Phi _0\left( t\right) \star G_0\left( x;t\mid x_0\right),
\end{equation}
with

\begin{equation}
\label{e1.6}
\Phi _0\left( t\right) = e^{-\lambda t},
\end{equation}
the {\it sojourn} probability at any site in the lattice.

The conditional probability density for the first passage time, in turn, can be expressed in terms of the Green's function in the Laplace representation by Siegert's \cite{siege} formula 
\begin{equation}
F_0^L\left( x;u\mid x_0\right) = \dfrac{G^L_0\left( x;u\mid x_0\right) }{G^L_0\left( x;u\mid x\right) }
\end{equation}

\subsection{Evanescent continuous time random walk}

We include now the possibility of evanescence of a walker at any site in the lattice. We assume that evanescence is a process statistically independent of displacement with a time independent rate $\eta $. In the evanescent case, the waiting time density (WTD) modifies to

\begin{align}
\label{e1b.1}
\psi _e\left( x-x\pri; t\right) = &\left[ p_d\, \delta _{x,x\pri +1} + p_i\, \delta _{x,x\pri -1} \right] \notag\\
& \times \lambda e^{-\left( \lambda +\eta \right) t}.
\end{align}
The Green's function for the evanescent CTRW (ECTRW) satisfies a recursive relation similar to Eq. (\ref{e1.2}),  where $\psi _0$ must be replaced by $\psi _e$. The solution is directly obtained in the Laplace representation in terms of $G_0^L$

\begin{equation}
\label{e1b.2}
G^L_e\left( x;u\mid x_0\right)  = G^L_0\left( x;u+\eta \mid x_0\right),  
\end{equation}
{\it i.e.}, in the standard CTRW Green's function in Eq. (\ref{e1.3}); the variable $u$ must be substituted by $u+\eta $.\\
If we assume an initially uniform concentration in the ECTRW, $c_0$, the concentration at time $t$ is 

\begin{equation}
\label{e1b.3}
c\left( t\right) = c_0 e^{-\eta t},
\end{equation}
{\it i.e.}, it remains uniform but decays exponentially.

\section{Local trap in an ECTRW}

We represent the reaction process as the trapping of a walker by a trap at a particular position in the lattice, denoted here as $x_1$. In Fig. \ref{fig1}, we represent the trapping as a transition of the walker to a limbo state from which it cannot return to the lattice. When a walker arrives at $x_1$, it may be trapped with a time independent rate $\kappa $ or it may make a transition to a neighbor site with a transition rate $\lambda $ continuing with its walk or it may even evanesce with rate $\eta $. We assume the three processes to be statistically independent from each other. To take into account the trapping possibility at $x_1$, the WTD at the trap position is modified to 

\begin{align}
\label{e2.1}
\psi _1\left( x-x_1; t\right) = &\left[ p_d\, \delta _{x,x_1 +1} + p_i\, \delta _{x,x_1 -1} \right] \notag\\
& \times \lambda e^{-\left( \lambda +\eta +\kappa \right) t}.
\end{align}
For the remaining sites in the lattice, the WTD is that of Eq. (\ref{e1b.1}). Therefore, the Green's function for the trapping problem, $G_t\left( x;t\mid x_0\right) $, satisfies the recursive relation

\begin{align}
\label{e2.1b}
G_t\left( x;t\mid x_0\right) = &\delta _{x,x_0} \delta \left( t-0^+\right) \\
& + \sum _{x\pri } \psi _i\left( x-x\pri; t\right) G_t\left( x\pri ;t\mid x_0\right),  \notag
\end{align}
with

\begin{equation}
\label{e2.1c}
\psi _i\left( x-x\pri; t\right) = \left\{ 
\begin{array}{ll}
\psi _e\left( x-x\pri; t\right) & x\pri \neq x_1 \\
 & \\
\psi _1\left( x-x_1; t\right) & x\pri = x_1
\end{array}
\right.
\end{equation}
By means of the local inhomogeneity method as in \cite{rebud,cacer}, we may express $G_t\left( x;t\mid x_0\right) $ in terms of $G_e\left( x;t\mid x_0\right) $ in the Laplace representation

\begin{align}
\label{e2.2}
G_t^L\left( x;u\mid x_0\right) = G_e^L\left( x;u\mid x_0\right) - \frac{A}{B}C,
\end{align}
where

\begin{align*}
A=& G_e^L\left( x;u\mid x_1\right) - \delta _{x,x_1} \\
&- \sum \limits _{x\pri }G_e^L\left( x;u\mid x\pri \right) \psi _1\left( x\pri - x_1\right), \\
B=&G_e^L\left( x_1;u\mid x_1\right) -\sum \limits_{x\pri }G_e^L\left( x_1;u\mid x\pri \right) \psi _1\left( x\pri - x_1\right), \\
C=&G_e^L\left( x_1;u\mid x_0\right).
\end{align*}

The Green's function in Eq. (\ref{e2.2}) at $x=x_1$ is of particular interest for the trapping problem. For a walker to be trapped, it must be at $x_1$ and it has to make a transition to the limbo state instead of making a transition to a neighbor site or to evanesce. The time of trapping probability density (APD) is given by the convolution product

\begin{equation}
\label{e2.3}
A\left( t\mid x_0\right) = \kappa e^{-\left( \kappa +\eta +\lambda \right) t}\star G^L_t\left( x_1;t\mid x_0\right). 
\end{equation}
An explicit expression for the APD is obtained in the Laplace representation by making use of Eqs. (\ref{e1.3}), (\ref{e1.4}), (\ref{e1b.2}) and (\ref{e2.2})

\begin{align}
\label{e2.4}
A^L\left( u\mid x_0\right) = &\dfrac{1}{1+2R_p\dfrac{\lambda }{\kappa }R\left( u+\eta \right) } \notag\\
&\times \dfrac{\left[ \xi \left( u+\eta \right) \right]^{\mid x_1-x_0\mid }}{R_c^{x_1-x_0}}.
\end{align}
The same expression for the APD is obtained if we assume that the trap (the minority species) is evanescent instead of the walkers (the majority species).

\section{Survival probability and reaction rate}

\begin{figure}[t]
\begin{center}
\includegraphics[width=0.98\columnwidth,trim=0 2cm 0 0,clip]{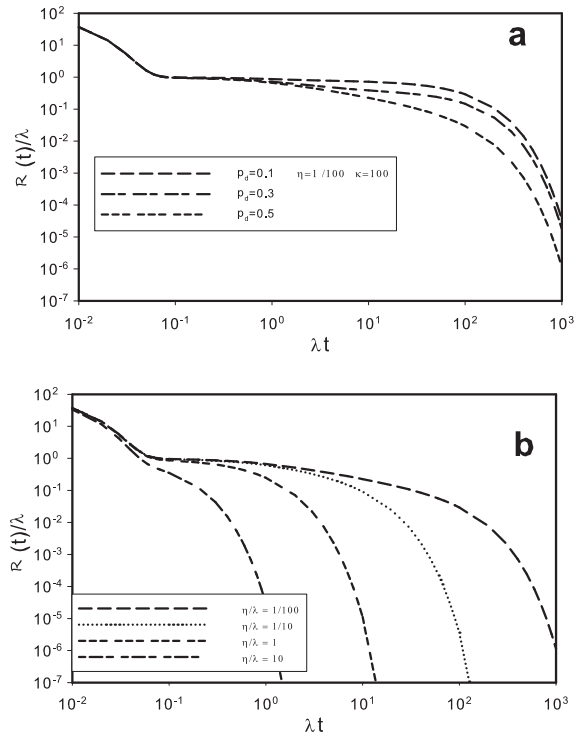}
\end{center}
\caption{Reaction rate as a function of time in dimensionless units. $\lambda $ is the transition rate between sites in the CTRW. Plot {\bf a} exhibits the influence of bias on the CTRW. Plot {\bf b} is for different values of the quotient $\eta /\lambda $, where $\eta $ is the evanescence rate.}
\label{fig2}
\end{figure}

If we consider a walker that starts its journey at $x_0$, the probability that this walker has not been trapped by time $t$ (one particle survival probability) is 

\begin{equation}
\label{e3.1}
S_1\left( t\right) = 1 - \int _0^t\, dt\pri \, A\left( t\pri \mid x_0\right).
\end{equation}
Following Bendler and Shlesinger \cite{bendl}, we assume a finite lattice of size $V$ with $N$ walkers initially uniformly distributed (the probability of finding a walker at a given site is $1/V$), with initial concentration $c_0=N/V$. The probability that none of the walkers initially on the lattice has been trapped by time $t$ is 

\begin{equation}
\label{e3.2}
S_N\left( t\right) = \left[ 1 - \frac{1}{V}\int _0^t\, dt\pri \, \sum _{x_0} A\left( t\pri \mid x_0\right) \right] ^N.
\end{equation}
In the thermodynamic limit ($N,V\rightarrow \infty$, $N/V\rightarrow c_0$), the probability of no reaction (survival probability) at time $t$ is 

\begin{equation}
\label{e3.3}
\Phi \left( t\right) = \exp \left[ -c_0\int_0^t\, dt\pri \, \sum _{x_0} A\left( t\pri \mid x_0\right) \right]. 
\end{equation}
The exponent in Eq. (\ref{e3.3}) is the integral of the time dependent reaction rate, ${\cal R} \left( t\right) = \partial _t \ln \left( \Phi \left( t\right) \right) $

\begin{equation}
\label{e3.4}
{\cal R} \left( t\right) = c_0\sum _{x_0} A\left( t\pri \mid x_0\right). 
\end{equation}
In Fig. \ref{fig2}, we present the reaction rate, ${\cal R} \left( t\right) $, obtained for the model in Eq. (\ref{e1.1}). The plots are presented in dimensionless units. As it can be appreciated in plot {\bf a}, there is a small influence of bias in ${\cal R}$ at intermediate times (in units of mean waiting time for diffusion). In plot {\bf b}, the effect of evanescence is shown. A faster decline in ${\cal R} $ is observed with increasing evanescence rate, as it should be expected.\\

\begin{figure}[t]
\begin{center}
\includegraphics[width=0.98\columnwidth,trim=0 2cm 0 0,clip]{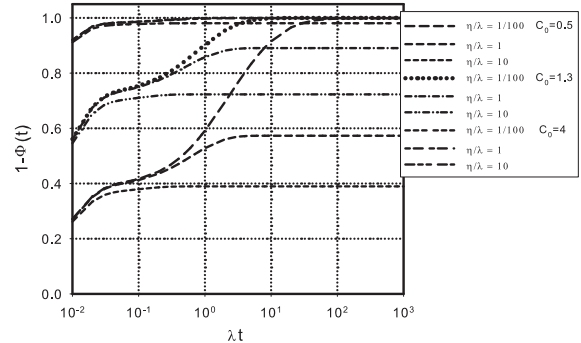}
\end{center}
\caption{$F\left( t\right) $, the fraction of minority species that have reacted by time $t$ vs. time in dimensionless units. $\lambda $ is the transition rate between sites in CTRW. For {\it immortal} walkers, the asymptotic limit is $1$. In this case, this value is reduced by evanescence.}
\label{fig3}
\end{figure}

In Fig. \ref{fig3}, we present the graph of the function $F\left( t\right) = 1-\Phi \left( t\right) $ vs. $\lambda t$ (dimensionless units). The function $F\left( t\right) $ introduced in \cite{chuan}, and also considered in \cite{rebud}, may be interpreted as the fraction of the original number of minority species (the trap) that have reacted by time $t$. At short times, the curves are grouped according to the value of the $\eta /\lambda $ quotient (we are considering only first order Markovian dynamics here), but at long times, this behavior is not observed due to evanescence.\\

\section{Discussion and conclusions}

We have presented a theoretical study of diffusion mediated reactions in an evanescent CTRW on a one dimensional lattice. A finite trapping rate is assumed when a walker reaches the trap position (imperfect trap model in Refs. \cite{colli,conda,rebud}). Therefore, in each encounter the reaction will not always occur and the walker may escape from the trap. Exact analytical expressions for the reaction rate and the survival probability (probability of no reaction) are obtained in the Laplace representation.\\
Evanescence modifies the long time behavior of the survival probability: the survival probability does not go to zero at long times as in the usual models. Consequently, the minority species fraction that reacts does not reach the asymptotic value of 1 at long times. In this case, the asymptotic value depends on the initial concentration of the majority species and on the ratio $\eta /\lambda $ between the evanescence rate and the diffusion rate.\\
Finally, we point out that the reaction rate value obtained is not modified if we assume an evanescent trap in the presence of non evanescent walkers.\\
More work along this line, generalizing the present results, is being developed and will be communicated elsewhere.\\

\end{document}